\documentclass[aps,amsfonts,amsmath,prd,preprint,nofootinbib,tightenlines]{revtex4}
\usepackage{epsf}

\newcommand{\beq}{\begin{equation}}
\newcommand{\eeq}{\end{equation}}

\def\be{\begin{equation}}
\def\ee{\end{equation}}
\def\beqa{\begin{eqnarray}}
\def\eeqa{\end{eqnarray}}

\begin{document}

\title{Probabilities in the Bousso-Polchinski multiverse}
\author{Delia Schwartz-Perlov and Alexander
Vilenkin}
\affiliation{Institute of Cosmology, Department of Physics and
Astronomy\\
Tufts University, Medford, MA 02155, USA }

\begin{abstract}
Using the recently introduced method to calculate bubble
abundances in an eternally inflating spacetime, we investigate the
volume distribution for the cosmological constant $\Lambda$ in the
context of the Bousso-Polchinski landscape model.  We find that the
resulting distribution has a staggered appearance which is in sharp
contrast to the heuristically expected flat distribution.  Previous
successful predictions for the observed value of $\Lambda$ have
hinged on the assumption of a flat volume distribution.  To
reconcile our staggered distribution with observations for
$\Lambda$, the BP model would have to produce a huge number of vacua
in the anthropic range $\Delta\Lambda_A$ of $\Lambda$, so that the
distribution could conceivably become smooth after averaging over
some suitable scale $\delta\Lambda\ll\Delta\Lambda_A$.


\end{abstract}
\maketitle

\section{Introduction}

The cosmological constant problem is one of the most intriguing
mysteries that we now face in theoretical physics. The observed
value of the cosmological constant $\Lambda$ is many orders of
magnitude smaller than theoretical expectations and is surprisingly
close to the present matter density of the universe,\footnote{Here
and below we use reduced Planck units, $M_p^2/8\pi =1$, where $M_p$
is the Planck mass.} \beq \Lambda_0\sim\rho_{m0}\sim 10^{-120}.
\label{Lambdarho} \eeq As of now, the only plausible explanation for
these enigmatic facts has been given in terms of the multiverse
picture, which assumes that $\Lambda$ is a variable parameter taking
different values in different parts of the universe
\cite{Weinberg87,Linde87,AV95,Efstathiou,MSW,GLV,Bludman,AV05}. The
probability for a randomly picked observer to measure a given value
of $\Lambda$ can then be expressed as \cite{AV95} \be
P_{obs}(\Lambda)\propto P(\Lambda)n_{obs}(\Lambda), \label{Pobs} \ee
where $P(\Lambda)$ has the meaning of the volume fraction of the
regions with a given value of $\Lambda$ and $n_{obs}(\Lambda)$ is
the number of observers per unit volume.\footnote{$P(\Lambda)$ is
often called the prior probability. Here we avoid this terminology,
since it is usually used to characterize one's ignorance or
prejudice, while the volume factor $P(\Lambda)$ should be
calculable, at least in principle.}  Disregarding the possible
variation of other ``constants'' and assuming that the density of
observers is roughly proportional to the fraction of matter
clustered in large galaxies, \be n_{obs}(\Lambda) \propto
f_G(\Lambda), \label{nobs} \ee one finds that the function
$n_{obs}(\Lambda)$ is narrowly peaked around $\Lambda =0$, with a
width
\beq \Delta\Lambda_A\sim 100\Lambda_0 \sim 10^{-118}.
\label{DeltaLambdaA} \eeq

In general, the volume factor $P(\Lambda)$ depends on the unknown
details of the fundamental theory and on the dynamics of eternal
inflation.  However, it has been argued \cite{AV96,Weinberg96} that
it should be well approximated by a flat distribution,
\be
P(\Lambda)\approx {\rm const}.
\label{flat}
\ee
The reason is that the anthropic range (\ref{DeltaLambdaA}), where the
function $n_{obs}(\Lambda)$ is substantially different from zero, is
much narrower than the full range of variation of $\Lambda$, which is
expected to be set by the Planck scale.  A smooth function varying on
this large characteristic scale will be nearly constant within the
tiny anthropic interval.

Combination of Eqs.~(\ref{Pobs})-(\ref{flat}) yields the
distribution \be P_{obs}(\Lambda)\propto f_G(\Lambda), \label{PnG}
\ee which can be readily calculated using the Press-Schechter
approximation for $f_G$. The observed value of $\Lambda$ is within
the $2\sigma$ range of this distribution -- an impressive success of
the multiverse paradigm.  One should keep in mind, however, that the
successful prediction for $\Lambda$ hinges on the assumption of a
flat volume distribution (\ref{flat}).  We emphasize that the form
of the volume distribution is important. If, for example, one uses
$P(\Lambda) \propto\Lambda$ instead of (\ref{flat}), the $2\sigma$
prediction would be $10 <\Lambda/\Lambda_0 <500$ and the observed
value of $\Lambda$ would be ruled out at a 99.9\% confidence level
\cite{Pogosian}.  The heuristic argument for a flat distribution
(\ref{flat}) sounds plausible, but it needs to be verified in
specific models.

The simplest model with a variable effective cosmological constant is
that of a scalar field $\phi$ with a very slowly varying potential
$V(\phi)$ \cite{Banks84,Linde87}. In such models, $\Lambda$ takes a
continuum range of values. It has been verified that the resulting
volume distribution for $\Lambda$ is indeed flat for a wide range of
potentials \cite{Weinberg00,GV00,GV03}. The main challenge one has to face
in this type of model is to justify the exceedingly flat potential
which is required to keep the field $\phi$ from rolling downhill on
the present Hubble time scale.

A model with a discrete spectrum of $\Lambda$ was first suggested by
Abbott \cite{Abbott}. He considered a scalar field with a
``washboard'' potential, having many local minima separated by
barriers. Transitions between the minima could occur through bubble
nucleation.  An alternative discrete model, first introduced by Brown
and Teitelboim \cite{BT}, assumes that the cosmological constant can
be expressed as
\be
\Lambda = \Lambda_{bare} +F^2/2.
\label{BT}
\ee
Here, $\Lambda_{bare}$ is the bare cosmological constant, which is
assumed to be large and negative, and $F$ is the magnitude of a
four-form field, which can change its value through the nucleation of
branes. The change of the field strength across the brane is
\be
\Delta F =\pm q,
\label{DeltaF}
\ee
where the ``charge'' $q$ is a constant fixed by the model.

In order to explain observations, the discrete spectrum of $\Lambda$
has to be very dense, with separation between adjacent values
\beq
\Delta\Lambda\lesssim \Lambda_0,
\label{dense}
\eeq
which in turn requires that the charge $q$ has to be very small. If
this is satisfied, analysis shows that the flat volume distribution
(\ref{flat}) is quite generic \cite{GV01}. But once again, the
exceedingly small charge $q$ required by the model appears to be
unnatural.\footnote{Some ideas on how such small parameters could
arise in particle physics have been suggested in
\cite{Weinberg00,Donoghue00,FMSW,BDM,DV01}.}

In an effort to remedy this problem, Bousso and Polchinski (hereafter
BP) extended the Brown-Teitelboim approach to include multiple
four-form fluxes \cite{BP}. They considered a model in which $J$
different fluxes give rise to a J-dimensional grid of vacua, each
labeled by a set of integers $n_a$. Each point in the grid corresponds
to a vacuum with the flux values $F_a = n_a q_a$ and a
cosmological constant
\be
\label{totalLambda}
\Lambda=\Lambda_{bare}+\frac{1}{2}\sum_{a=1}^{J}F_a^2=\Lambda_{bare}
+\frac{1}{2}\sum_{a=1}^{J}n_a^2q_a^2.
\ee
This model is particularly interesting because multiple fluxes
generally arise in string theory compactifications. The model can thus
be regarded as a toy model of the string theory landscape.
BP showed that with $J\sim 100$, the spectrum of allowed values of
$\Lambda$ can be sufficiently dense, even in the absence of very small
parameters, e.g., with $|\Lambda_{bare}|\sim 1$, $q_a\sim 0.1$.

In the cosmological context, high-energy vacua of the BP grid will
drive exponential inflationary expansion. The flux configuration in
the inflating region can change from one point on the grid to the next
through bubble nucleation. Bubbles thus nucleate within bubbles, and
each time this happens the cosmological constant either increases or
decreases by a discrete amount. This mechanism allows the universe to
start off with an arbitrary large cosmological constant, and then to
diffuse through the BP grid of possible vacua, to generate regions
with each and every possible cosmological constant, including that
which we inhabit.

Our goal in this paper is to study the volume distribution for
$\Lambda$ in the BP model. In particular, we would like to check
whether or not this distribution is approximately flat, as suggested
by the heuristic argument of \cite{AV96,Weinberg96}. Until recently,
such an analysis would have been rather problematic, since the
calculation of the volume fractions in an eternally inflating
universe is notoriously ambiguous. The volume of each type of vacuum
diverges in the limit $t\to\infty$, so in order to calculate
probabilities, one has to impose some kind of a cutoff. The answer,
however, turns out to be rather sensitive to the choice of the
cutoff procedure. If, for example, the cutoff is imposed on a
constant time surface, one gets very different distributions
depending on one's choice of the time variable $t$ \cite{LLM}. (For
more recent discussions, see \cite{VVW,Guth00,Tegmark}.)

Fortunately, a fully gauge-invariant prescription for calculating
probabilities has been recently introduced in \cite{GSPVW}.
It has been tried on some simple models and seems to give reasonable
results. Here we shall apply it to the BP model.

As BP themselves recognized, their model does not give an accurate
quantitative description of the string theory landscape. In
particular, it does not explain how the sizes of compact dimensions
get stabilized. This issue was later addressed by Kachru, Kallosh,
Linde and Trivedi (KKLT) \cite{KKLT}, who provided the first example
of a metastable string theory vacuum with a positive cosmological
constant. Apart from the flux contributions in (\ref{totalLambda}),
the vacuum energy in KKLT-type vacua gets contributions from
non-perturbative moduli potentials and from branes.  The $4D$
Newton's constant in these vacua depends on the volume of extra
dimensions and changes from one vacuum to another.  Douglas and
collaborators \cite{Douglas,AshokDouglas,DenefDouglas} studied the
statistics of KKLT-type vacua. Their aim was to find the number of
vacua with given properties (e.g., with a given value of $\Lambda$)
in the landscape. Our goal here is more ambitions: we want to find
the probability for a given $\Lambda$ to be observed.

In our analysis, we shall disregard all the complications of the
KKLT vacua, with the hope that the simple BP model captures some of
the essential features of the landscape. At the end of the paper we
shall discuss which aspects of our results can be expected to
survive in more realistic models.

We begin in the next section by summarizing the prescription of
Ref.~\cite{GSPVW} for calculating probabilities.  We shall see that
the problem reduces to finding the smallest eigenvalue and the
corresponding eigenvector of a large matrix, whose matrix elements
are proportional to the transition rates between different vacua.
The calculation of the transition rates for the BP model is reviewed
in Section III.  Some of these rates are extremely small, since the
upward transitions with an increase of $\Lambda$ are exponentially
suppressed relative to the downward transitions. In Section IV we
develop a perturbative method for solving our eigenvalue problem,
using the upward transition rates as small parameters. This method
is applied to the BP model in Section V. We find that the resulting
probability distribution has a very irregular, staggered appearance
and is very different from the flat distribution (\ref{flat}). The
implications of our results for the string theory landscape are
discussed in Section VI.

\section{Prescription for probabilities}


Here we summarize the prescription for calculating the volume
distribution proposed in Ref.~\cite{GSPVW}. Suppose we have a theory
with a discrete set of vacua, labeled by index $j$. The cosmological
constants $\Lambda_j$ can be positive, negative, or zero. Transitions
between the vacua can occur through bubble nucleation. The proposal of
Ref.~\cite{GSPVW} is that the volume distribution is given by
\be
P_j \propto p_j Z_j^3,
\label{PpZ}
\ee
where $p_j$ is the relative abundance of bubbles of type $j$ and $Z_j$
is (roughly) the amount of slow-roll inflationary expansion inside the
bubble after nucleation (so that $Z_j^3$ is the volume slow-roll
expansion factor).

The total number of nucleated bubbles of any kind in an eternally
inflating universe is known to grow without bound, even in a region
of finite comoving size. We thus need to cut off our count. The
proposal of \cite{GSPVW} is that the counting should be done at the
future boundary of spacetime and should include only bubbles with
radii greater than some tiny co-moving size $\epsilon$. The limit
$\epsilon\rightarrow 0$ should then be taken at the end. (An
equivalent method for calculating $p_j$ was suggested in
\cite{ELM}.) It was shown in \cite{GSPVW} that this prescription is
insensitive to the choice of the time coordinate and to coordinate
transformations at future infinity.

The bubble abundances $p_j$ can be related to the functions $f_j(t)$
expressing the fraction of comoving volume occupied by vacuum $j$ at
time $t$. These functions obey the evolution equation \cite{recycling}
\beq
{df_j\over{dt}}=\sum_i (-\kappa_{ij}f_j + \kappa_{ji}f_i),
\label{dfdt}
\eeq
where the first term on the right-hand side accounts for loss of
comoving volume due to bubbles of type $i$ nucleating within those of
type $j$, and the second term reflects the increase of comoving volume
due to nucleation of type-$j$ bubbles within type-$i$ bubbles.

The transition rate $\kappa_{ij}$ is defined as the probability per
unit time for an observer who is currently in vacuum $j$ to find
herself in vacuum $i$. Its magnitude depends on the choice of the time
variable $t$. The most convenient choice for our purposes is to use
the logarithm of the scale factor as the time variable; this is the
so-called scale-factor time. With this choice,
\beq
\kappa_{ij}=\Gamma_{ij}{4\pi\over{3}}H_j^{-4},
\label{kappa}
\eeq
where $\Gamma_{ij}$ is the bubble nucleation rate per unit
physical spacetime volume (same as $\lambda_{ij}$ in \cite{GSPVW})
and
\be
H_j = (\Lambda_j/3)^{1/2}
\label{Hj}
\ee
is the expansion rate in vacuum $j$.

We distinguish between the recyclable, non-terminal vacua, with
$\Lambda_j>0$, and the non-recyclable, "terminal vacua", for which
$\Lambda_j\leq 0$.  Transitions from either a flat spacetime
($\Lambda_j=0$), or a negative $\Lambda$ FRW spacetime
($\Lambda_j<0$), which increase $\Lambda$ have a zero probability of
occurring.\footnote{This is because the volume of the instanton is
compact whilst the volume of the Euclideanized background spacetime
is infinite, so that the difference in their actions is infinite.}
Transitions from $\Lambda_j = 0$ and from small negative $\Lambda_j$
to even more negative $\Lambda$ are possible, but these $\Lambda$'s
are likely to be large and negative and are therefore of no
anthropic interest. We will label the recyclable, non-terminal vacua
by Greek letters, and for non-recyclable, terminal vacua, we will
reserve the indices $m$ and $n$.  Then, by definition, \beq
\Gamma_{\alpha m}=\Gamma_{mn}=0. \label{lambda=0} \eeq Latin letters
other than $m,n$ will be used to label arbitrary vacua, both
recyclable and terminal, with the exception of letters $a,b$, which
we use to label the fluxes.

Eq.~(\ref{dfdt}) can be written in a vector form,
\beq
{d{\bf f}\over{dt}}={\mathbf M}{\bf f},
\label{matrixf}
\eeq
where ${\bf f(t)}\equiv \{ f_j(t)\}$ and
\beq
M_{ij}=\kappa_{ij}-\delta_{ij}\sum_r \kappa_{ri}.
\label{Mij}
\eeq

The asymptotic solution of (\ref{matrixf}) at large $t$ has the
form
\beq
{\bf f}(t)={\bf f^{(0)}}+{\bf s}e^{-q t}+ ...
\label{asympt}
\eeq
Here, ${\bf f}^{(0)}$ is a constant vector
which has nonzero components only in terminal vacua,
\beq
f_\alpha^{(0)}=0,
\label{falpha=0}
\eeq
while the values of $f_n^{(0)}$ depend on the choice of initial
conditions.  It is clear from Eq.~(\ref{lambda=0}) that any such
vector is an eigenvector of the matrix ${\mathbf M}$ with zero
eigenvalue, \beq {\mathbf M}{\bf f_0}=0. \eeq As shown in
\cite{GSPVW}, all other eigenvalues of $\mathbf{M}$ have a negative
real part, so the solution approaches a constant at late times. We
have denoted by $-q$ the eigenvalue with the smallest (by magnitude)
negative real part and by $\mathbf{s}$ the corresponding eigenvector.

It has been shown in \cite{GSPVW} that the bubble abundances $p_j$ can
be expressed in terms of the asymptotic solution (\ref{asympt}). The
resulting expression is
\beq
p_j\propto \sum_\alpha H_\alpha^q \kappa_{j\alpha}
s_\alpha.
\label{pJaume}
\eeq
where the summation is over all recyclable vacua which can directly
tunnel to $j$.

Note that the calculation of $p_j$ requires only knowledge of the
components $s_\alpha$ for the recyclable vacua. The evolution of the
comoving volume fraction in these vacua is independent of that in the
terminal vacua. Formally, this can be seen from the fact that the
transition matrix $\mathbf M$ in (\ref{matrixf}) has the form
 \be \mathbf{M}=
\left(%
\begin{array}{cc}
  \mathbf{R} & 0 \\
  \mathbf{T} & 0 \\
\end{array}%
\right) \ee Here, $\mathbf{R}$ is a square matrix with matrix
elements between recyclable vacua, while the matrix elements of
$\mathbf{T}$ correspond to transitions from recyclable vacua to
terminal ones. Eigenvectors of $\mathbf{M}$ are of the form ${\bf
f}=(\bf s,\bf t)$, where $s$ is an eigenvector of ${\mathbf R}$,
\beq \mathbf{R}{\bf s}=-q{\bf s}, \label{Rs} \eeq and ${\bf t}$ is
arbitrary. Eigenvalues of  $\mathbf{M}$ are the same as those of
$\mathbf{R}$, except for some additional zero eigenvalues with
eigenvectors which only have nonzero entries for terminal vacua.

The problem of calculating $p_j$ has thus been reduced to finding the
dominant eigenvalue $q$ and the corresponding eigenvector ${\bf s}$ of
the recyclable transition matrix $\mathbf{R}$. In the following
sections we shall apply this prescription to the BP model.

\section{Nucleation rates in the BP model}

In the BP model, we have a $J$-dimensional grid of vacua characterized
by the fluxes $F_a=n_a q_a$ and vacuum energy densities given by
Eq. (\ref{totalLambda}). BP emphasized that $q_a$ need not be very
small, $q_a/|\Lambda_{bare}|\sim 0.1~-~1$. So the model does not have
any small parameters, except perhaps the values of $\Lambda_j$ in some
vacua, where the contribution of fluxes is nearly balanced by
$\Lambda_{bare}$.

Transitions between the neighboring vacua, which change one of the
integers $n_a$ by $\pm 1$ can occur through bubble
nucleation. The
bubbles are bounded by thin branes, whose tension $\tau_a$ is related
to their charge $q_a$ as
\beq
\tau_a^2 =q_a^2/2.
\label{tauj}
\eeq
The latter relation is suggested by string theory
\cite{BP,FMSW}. It applies only in the supersymmetric limit, but
we shall neglect possible corrections due to supersymmetry
breaking. Transitions with multiple brane nucleation, in
which $|\Delta n_a|>1$ or several $n_a$ are changed at once, are
likely to be strongly suppressed \cite{Megevand}, and we shall
disregard them here.

The bubble nucleation rate $\Gamma_{ij}$ per unit spacetime volume can
be expressed as \cite{CdL}
\be
\Gamma_{ij}=A_{ij} \exp^{-B_{ij}}
\label{Gamma}
\ee
with
\beq
B_{ij}=I_{ij}-S_j
\label{Bij}
\eeq
Here, $I_{ij}$ is the Coleman-DeLuccia instanton action and
\beq
S_j=-{8\pi^2\over{H_j^2}}
\label{Sj}
\eeq
is the background Euclidean action of de Sitter space.

In the case of a thin-wall bubble, which is appropriate for our
problem, the instanton action $I_{ij}$ has been calculated in
Refs.~\cite{CdL,BT}. It depends on the values of $\Lambda$ inside
and outside the bubble and on the brane tension $\tau$.

Let us first consider a bubble which changes the flux $a$ from $n_a$
to $n_a-1$ ($n_a>0$).  The resulting change in the cosmological constant is
given by
\be
|\Delta\Lambda_a|=(n_a-1/2)q_a^2,
\label{DeltaLambda}
\ee
and the exponent in the tunneling rate (\ref{Gamma}) can be
expressed as
\be
B_{a\downarrow} = B_{a\downarrow}^{flatspace} r(x,y).
\label{Bdown}
\ee
Here, $B_{a\downarrow}^{flatspace}$ is the flat space
bounce action,
\be
B_{a\downarrow}^{flatspace}= \frac{27
\pi^2}{2}\frac{\tau_a^4}{|\Delta \Lambda_a|^3}.
\ee
With the aid of
Eqs.~(\ref{tauj}),(\ref{DeltaLambda}) it can be expressed as
\be
B_{a\downarrow}^{flatspace}= \frac{27
\pi^2}{8}\frac{1}{(n_a-1/2)^3q_a^2}
\label{Bflat}
\ee

The gravitational correction factor $r(x,y)$ is given by
\cite{Parke}
\be
r(x,y) = \frac{2[(1+x
y)-(1+2xy+x^2)^{\frac{1}{2}}]}{x^2(y^2-1)(1+2xy+x^2)^{\frac{1}{2}}}
\label{gravfactor}
\ee
with the dimensionless parameters
\be
x\equiv
\frac{3q_a^2}{8|\Delta\Lambda_a|}=\frac{3}{8(n_a-1/2)}
\ee
and
\be
y\equiv \frac{2\Lambda}{|\Delta\Lambda_a|}-1, \label{y} \ee where
$\Lambda$ is the background value prior to nucleation.

The prefactors $A_{ij}$ in (\ref{Gamma}) can be estimated as
\beq
A_{ij} \sim 1.
\label{Aij}
\eeq
This is an obvious guess for nucleation out of vacua with
$\Lambda_j\sim 1$. (This guess is supported by the detailed analysis
in Ref.~\cite{Jaume}.)  For $\Lambda_j \ll 1$, we still expect
Eq.(\ref{Aij}) to hold, since we know that the tunneling rate remains
finite in the limit $\Lambda_j \to 0$, $|\Delta\Lambda_a|\sim 1$.

If the vacuum $n_a-1$ still has a positive energy density, then an
upward transition from $n_a -1$ to $n_a$ is also possible. The
corresponding transition rate is characterized by the same instanton
action and the same prefactor \cite{EWeinberg}, \beq I_{ij}=I_{ji},
~~~~~~~ A_{ij}=A_{ji}, \label{ijji} \eeq and it follows from
Eqs.(\ref{Gamma}), (\ref{Bij}) and (\ref{Hj}) that the upward and
downward nucleation rates are related by \be \Gamma_{ji} =
\Gamma_{ij} \exp\left[24 \pi^2
\left(\frac{1}{\Lambda_{j}}-\frac{1}{\Lambda_{i}}\right)\right].
\label{updown} \ee The exponential factor on the right-hand side of
(\ref{updown}) depends very strongly on the value of $\Lambda_{j}$.
The closer we are to $\Lambda_j=0$, the more suppressed are the
upward transitions $j\to i$ relative to the downward ones.

\begin{figure}
\begin{center}
\leavevmode\epsfxsize=5in\epsfbox{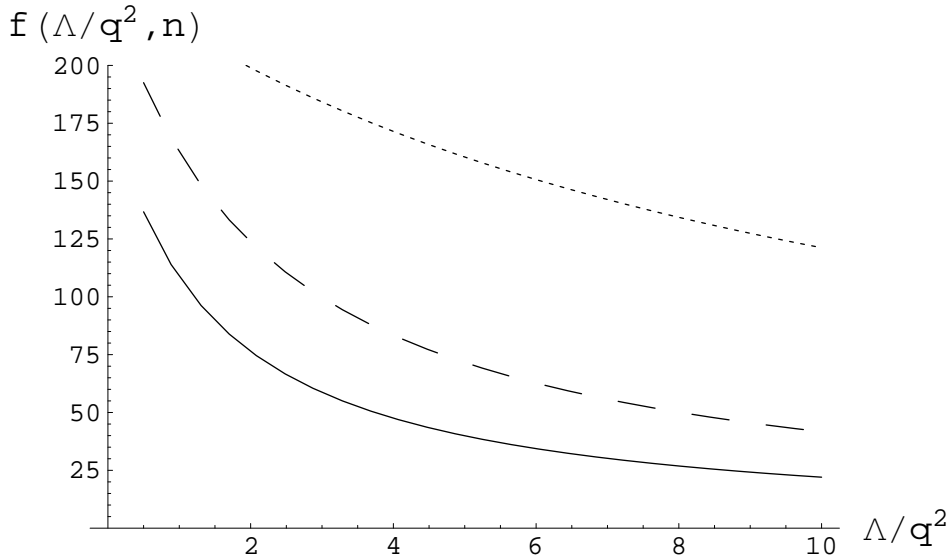}
\end{center}
\caption{The factor $f(\Lambda/q_a^2,n_a)$ as a function of
$\Lambda/q_a^2$ for $n_a=1$ (solid line), $n_a=2$ (dashed line), and
$n_a=10$ (dotted line).} \label{7Dgammadownuplam}
\end{figure}

Eq.~(\ref{updown}) shows that the transition rate from $n_a$ up to
$n_{a+1}$ is suppressed relative to that from $n_{a+1}$ down to
$n_a$. It can also be shown that upward transitions from $n_a$ to
$n_{a+1}$ are similarly suppressed relative to the downward
transitions from $n_a$ to $n_{a-1}$.  Using
Eqs.~(\ref{DeltaLambda})-(\ref{y}), the ratio of the corresponding
rates can be expressed as \beq
\ln(\Gamma_{\downarrow}/\Gamma_{\uparrow}) =
\Lambda^{-1}f(\Lambda/q_a^2,n_a). \label{Gammaf} \eeq The factor
$f(\Lambda/q_a^2,n_a)$ is plotted in Fig. \ref{7Dgammadownuplam} as
a function of $\Lambda/q_a^2$ for $n_a=1$ and $n_a=2$.  The plot
shows that upward transitions are strongly suppressed, unless
$\Lambda/q_a^2$ is very large. The factor $\Lambda^{-1}$ in
Eq.~(\ref{Gammaf}) results in even stronger suppression when
$\Lambda$ is well below the Planck scale.


Transition rates from a given vacuum $j$ to different states $i$ are
related by
\beq
\Gamma_{ij}\propto \exp(-I_{ij}).
\label{GammaI}
\eeq
As a rule of thumb,
\beq
I_{ij}\sim -\Lambda_{max}^{-1},
\label{ILambda}
\eeq
where $\Lambda_{max}$ is the larger of $\Lambda_i$ and $\Lambda_j$. It
follows from (\ref{GammaI}),(\ref{ILambda}) that upward transitions
from a given site are more probable to the lower-energy vacua.

To develop some intuition for the dependence of the tunneling
exponent $B_{a\downarrow}$ on the parameters of the model, we shall
consider the limits of small and large $\Lambda$. For $\Lambda\ll
|\Delta\Lambda_a|$, we have $y\approx -1$, and Eq.
(\ref{gravfactor}) gives \beq r(y\to -1)=(1-x)^{-2} >1. \eeq Hence,
for low-energy vacua the tunneling exponent is increased over its
flat-space value, resulting in a suppression of the nucleation rate.
(For small values of $x$, $r$ is increased only by a small fraction,
but the factor $B_{a\downarrow}^{flatspace}$ that it multiplies in
the tunneling exponent is typically rather large, so the resulting
suppression can still be significant.)

In the opposite limit, $\Lambda\gg|\Delta\Lambda_a|$, $y\gg 1$, \beq
r(y\gg 1)\approx \sqrt{2}(xy)^{-3/2}, \eeq and \beq
B_{a\downarrow}\approx \frac{27\pi^2}{2} q_a
\left(\frac{2}{3\Lambda}\right)^{3/2}. \label{LargeLambda} \eeq

\begin{figure}
\begin{center}
\leavevmode\epsfxsize=5in\epsfbox{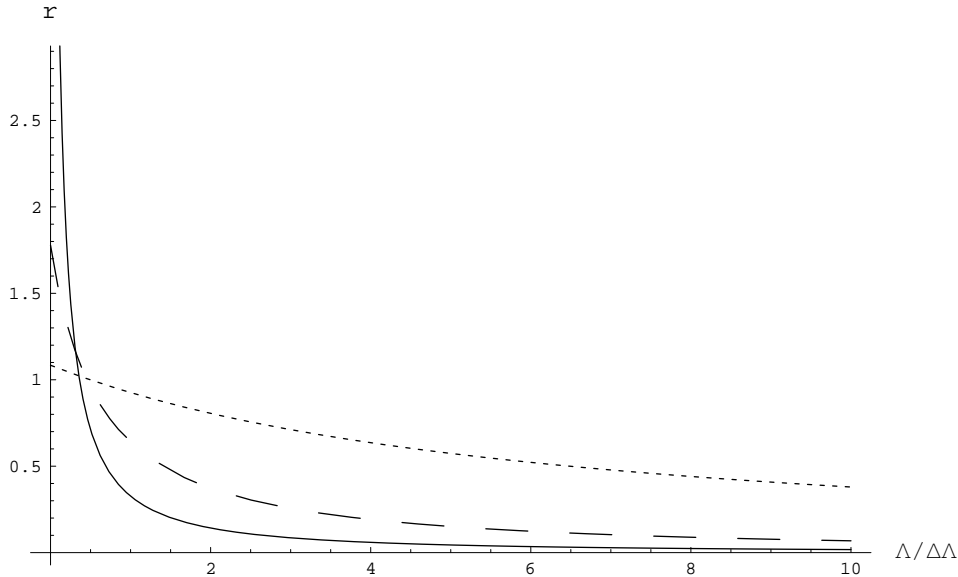}
\end{center}
\caption{The gravitational factor $r$ as a function of
$\Lambda/|\Delta\Lambda_a|$ for $n_a=1$ (solid line), $n_a=2$
(dashed line), and $n_a=10$ (dotted line).} \label{7Drvsz}
\end{figure}

The gravitational factor $r$ is plotted in Fig. \ref{7Drvsz}
as a function of $\Lambda/|\Delta\Lambda_a|$ for $n_a=1$, $n_a=2$ and
$n_a=10$ (corresponding to $x=3/4,~1/4$ and $0.04$, respectively). We
see that for large values of $\Lambda$, $r\ll 1$, so the nucleation
rate is enhanced.  In order for our model to be viable, we must ensure
that the tunneling action is large enough to justify the use of the
semi-classical approximation: $B_{a\downarrow}\gg 1$, or
$\Lambda/q_a^2\ll 20 q_a^{-4/3}$.

If $q_a$ and $\Lambda$ are changed simultaneously, keeping the
ratios $\Lambda/q_a^2$ fixed, then $x$ and $y$ do not change, and it
is clear from Eqs.~(\ref{Bdown}),(\ref{Bflat}),(\ref{gravfactor})
that the nucleation exponents scale as $B_{ij}\propto \Lambda^{-1}$.
This shows that bubble nucleation rates are strongly suppressed when
the energy scales of $q_a$ and $\Lambda$ are well below the Planck
scale.

\section{Perturbation theory}

\subsection{Degeneracy factors}

We shall assume for simplicity that the integers $n_a$ take values in
the range $|n_a|\leq N$, where $N$ is independent of $a$. The number
of vacua in the grid is then $(2N+1)^J$.

To maximize computational abilities, we used the symmetry $n_a\to
-n_a$ and restricted the analysis to the sector $0\leq\{n_a\}\leq
N$. We took into account the degeneracies in $\Lambda$ that would
occur if we allowed negative values of $n_a$ by assigning
appropriate degeneracy factors to the probabilities that we
calculated. For example, if we have a two-dimensional grid, $J=2$,
and only consider the quadrant $n_a\geq 0$, then any point that lies
on one of the two axes will be doubly degenerate (configuration
$\{0,1\}$ has the same $\Lambda$ as $\{0,-1\}$), whilst a point that
lies in the interior of the quadrant will have a four-fold
degeneracy (configuration $\{1,2\}$ has the same $\Lambda$ as
$\{-1,2\}$, $\{-1,-2\}$, and $\{1,-2\}$).

In general, the degeneracy of each site can be calculated
using the following formula:
\be
{\cal D}\{n_a\}= 2^{k\{n_a\}},
\ee
where
\beq
k\{n_a\} = J - (\delta_{0 n_1} + \delta_{0 n_2} +...+ \delta_{0
n_J})
\ee
So points which have no zero coordinates for a J=7 model have
${\cal D} = 2^7 =128$. A point with one zero coordinate has
${\cal D} = 2^6 =64$ etc.

When we use Eq. (\ref{pJaume}) to calculate the
probabilities, we multiply the RHS by the appropriate degeneracy.

Diffusion from a grid point for which $n =0$ to $n_a =-1$,
is equivalent to the diffusion from $n =0$ to $n_a =+1$.  Also,
diffusion from $n_a =-1$ to $n =0$ is equivalent to $n_a =1$ to $n
=0$.  Thus we were able to take into account these processes in the
transition matrix by double counting the positive $n_a$ to or from
$n=0$ transition rates.  In summary, we implemented boundary
conditions such that our process is equivalent to diffusion through
a J-dimensional grid, with $-N\leq n_a \leq N$.

\begin{figure}
\begin{center}
\leavevmode\epsfxsize=5in\epsfbox{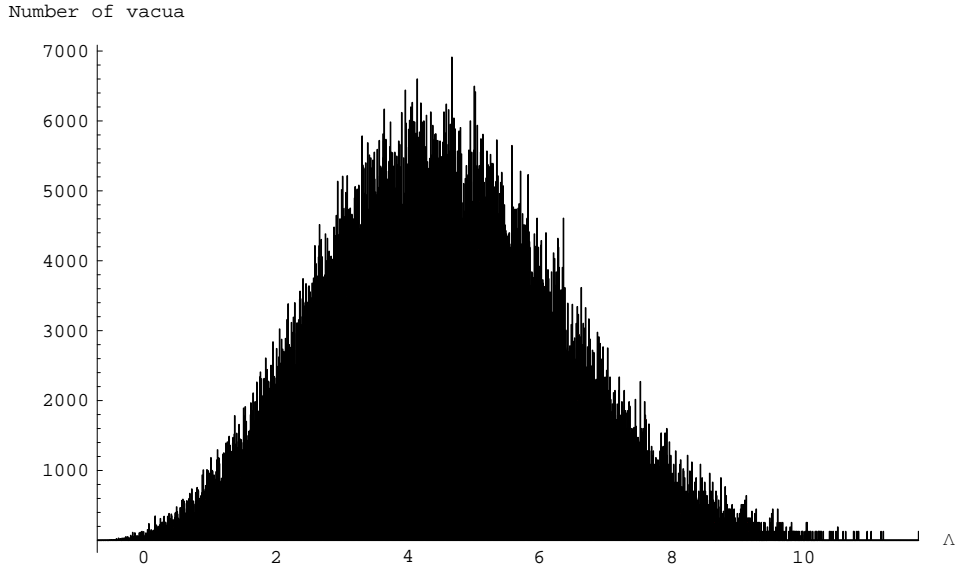}
\end{center}
\caption{The spectrum of vacua for a $J=7$, $N=4$ BP grid with
parameters given in (\ref{7D}).}
\label{7Dspectrumrough}
\end{figure}

\begin{figure}
\begin{center}
\leavevmode\epsfxsize=5in\epsfbox{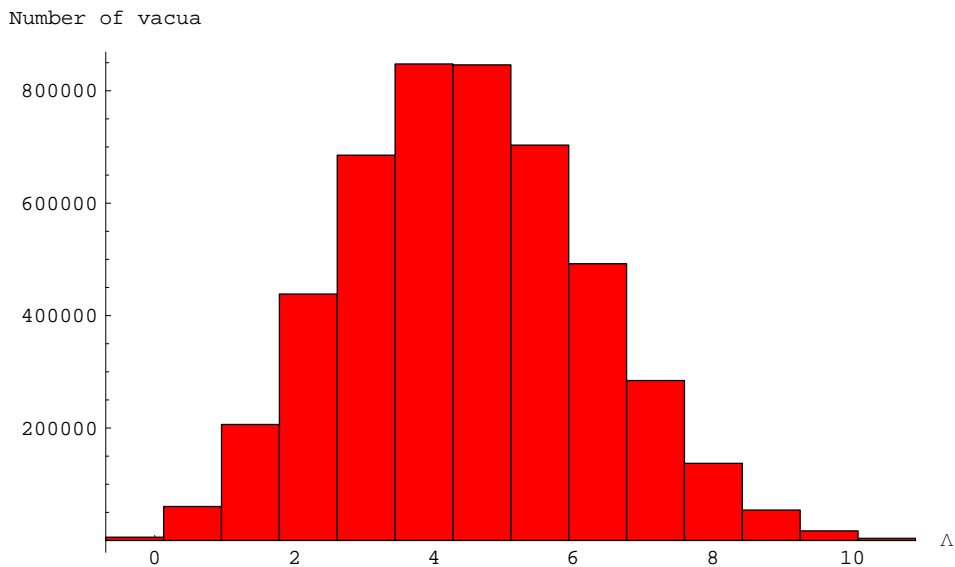}
\end{center}
\caption{The smoothed spectrum for the above model.}
\label{7Dspectrum}
\end{figure}

As an illustrative example, we show in Fig. \ref{7Dspectrumrough} a
histogram of the number of vacua vs. $\Lambda$ for a model with
$J=7$ and $N=4$, which has $\sim 10^7$ vacua. The parameter values
used in this model are \beq {q_a}= {0.5308,~0.3909,~0.5175,~0.4722,~
0.5103,~0.4036,~0.4541}; ~~~~~~~\Lambda_{bare}=-0.702. \label{7D}
\eeq The sharp spikes and dips in the histogram are due partly to
the non-uniform distribution of the vacua along the $\Lambda$-axis
and partly to the difference in degeneracy factors for different
vacua. The spikes disappear when the histogram is plotted with a
larger bin size, as shown in Fig.\ref{7Dspectrum}.

\subsection{Zeroth order}

As outlined in Section II, the calculation of probabilities reduces
to finding the smallest eigenvalue and the corresponding eigenvector
for a huge ${\cal N}\times {\cal N}$ (recycling) transition matrix
$\mathbf{R}$. Here, ${\cal N}$ is the number of recyclable vacua,
which we expect to be comparable to the total number of vacua. In
our numerical example ${\cal N}\sim 10^7$, while for a realistic
string theory landscape it can be as large as $10^{500}$
\cite{Susskind,Douglas,AshokDouglas,DenefDouglas}. Matters are
further complicated by the fact that some of the elements of
$\mathbf{R}$ are exceedingly small. For example, it follows from
Eq.~(\ref{updown}) that upward transitions from low-energy vacua
with $\Lambda_j\ll 1$ are very strongly suppressed. Matrix
diagonalization programs like Mathematica are not well suited for
dealing with such matrices.  We shall see, however, that the
smallness of the upward transition rates can be used to solve our
eigenvalue problem via perturbation theory, with the upward
transition rates playing the role of small expansion parameters.

We represent our transition matrix as a sum of an unperturbed matrix
and a small correction, \be \mathbf{R}=\mathbf{R^{(0)}}+
\mathbf{R^{(1)}}, \ee where $\mathbf{R^{(0)}}$ contains all the
downward transition rates and $\mathbf{R^{(1)}}$ contains all the
upward transition rates.  We will solve for the zero'th order
dominant eigensystem $\{q^{(0)},\mathbf{s^{(0)}}\}$ from
$\mathbf{R^{(0)}}$ and then find the first order corrections by
including contributions from $\mathbf{R^{(1)}}$. Note that the
eigenvalue correction $q^{(1)}$ is not needed for the calculation of
bubble abundances (\ref{pJaume}) to the lowest nonzero order. One
only needs to calculate the eigenvector correction
$\mathbf{s^{(1)}}$ (since the zeroth-order components
$s_\alpha^{(0)}$ vanish for recyclable vacua).


If the vacua are arranged in the order of increasing $\Lambda$, so
that \beq \Lambda_1 \leq \Lambda_2 \leq ... \leq \Lambda_{\cal N},
\eeq then $\mathbf{R^{(0)}}$ is a triangular matrix, with all matrix
elements below the diagonal equal to zero. Its eigenvalues are
simply equal to its diagonal elements, \beq R^{(0)}_{\alpha\alpha}
=-\sum_{j<\alpha}\kappa_{j\alpha} \equiv -D_\alpha.
\label{Ralpha}\eeq Hence, the magnitude of the smallest zeroth-order
eigenvalue is \beq q^{(0)}=D_{{\alpha_*}} \equiv {\rm
min}\{D_\alpha\}. \label{q0} \eeq

Up to the coefficient $(4\pi/3)H_\alpha^{-4}$, $D_\alpha$ is the total
decay rate of vacuum $\alpha$. As we discussed in Section III, bubble
nucleation rates are exponentially suppressed in low-energy vacua with
$\Lambda_j\ll 1$. We therefore expect that the vacuum ${\alpha_*}$
corresponding to the smallest eigenvalue $q^{(0)}$ is one of the
low-energy vacua.

With $\Lambda_{\alpha_*}\ll 1$ and $q_a$ not very small,
Eq.(\ref{DeltaLambda}) suggests that downward transitions from
${\alpha_*}$ will bring us to the negative-$\Lambda$ territory of
terminal vacua.  Terminal vacua do not belong in the matrix
$\mathbf{R}$; hence, $R_{\beta{\alpha_*}}=0$ for
$\beta\neq{\alpha_*}$, and it is easy to see that our zeroth order
eigenvector has a single nonzero component, \beq s^{(0)}_\alpha=
\delta_{\alpha{\alpha_*}}. \label{s0} \eeq Eq. (\ref{pJaume}) then
implies that the only vacua with nonzero probabilities at zeroth order
are the negative-$\Lambda$ descendants which can be reached by a
single downward jump from the dominant vacuum ${\alpha_*}$.  (Note
that the vacuum $\alpha_*$ itself has zero probability at this order.)

\subsection{First order}

The full eigenvalue equation can be written as
\beq
({\mathbf R^{(0)}}+{\mathbf R^{(1)}})({\bf s^{(0)}} + {\bf
s^{(1)}}) = -(q^{(0)}+q^{(1)}) ({\bf s^{(0)}} + {\bf
s^{(1)}}).
\eeq
Neglecting second-order terms and using the zeroth-order relation
\beq
{\mathbf R^{(0)}}{\bf s^{(0)}} = - q^{(0)}{\bf s^{(0)}},
\eeq
we obtain an equation for the first-order corrections,
\beq
({\mathbf R^{(0)}}+q^{(0)}{\mathbf I}){\bf s^{(1)}} =
-({\mathbf R^{(1)}}+q^{(1)}{\mathbf I}){\bf s^{(0)}},
\label{firstorder}
\eeq
where ${\mathbf I}$ is the unit matrix.

Eq. (\ref{firstorder}) is a system of ${\cal N}$ linear equations
for the ${\cal N}$ components of ${\bf s^{(1)}}$. Note, however,
that the triangular matrix multiplying ${\bf s^{(1)}}$ on the
left-hand side has a zero diagonal element, \beq ({\mathbf
R^{(0)}}+q^{(0)}{\mathbf I})_{{\alpha_*}{\alpha_*}} = 0, \eeq which
means that the determinant of this matrix vanishes, so it cannot be
inverted. In other words, the equations in (\ref{firstorder}) are
not all linearly independent.

This problem can be cured by dropping the ${\alpha_*}$-th equation
in (\ref{firstorder}) and replacing it by a constraint equation, which
we choose to enforce the orthogonality of ${\bf s^{(1)}}$ and ${\bf
s^{(0)}}$,
\beq
({\bf s^{(0)}},{\bf s^{(1)}}) =0.
\label{constraint}
\eeq
Note that the ${\alpha_*}$-th equation is the only equation in
(\ref{firstorder}) that involves the eigenvalue correction
$q^{(1)}$. Now $q^{(1)}$ has dropped out of our modified system, and
we can straightforwardly solve for $s^{(1)}_\alpha$. We did this
numerically for a $J=7$ model; the results will be presented in
the following subsection. A $J=2$ analytic toy model is worked out in
the Appendix.


\section{Bubble abundances in the BP model}

We found in the preceding section that the zeroth order of
perturbation theory picks the vacuum $\alpha_*$ which decays the
slowest (we call it the dominant vacuum), and assigns non-zero
probabilities to it's offspring only - all other probabilities are
zero. In the first order of perturbation theory, all vacua connected
to the dominant vacuum via one upward jump, and any vacua connected to
these via a series of downward transitions, also acquire non-zero
probabilities.

\begin{figure}
\begin{center}
\leavevmode\epsfxsize=5in\epsfbox{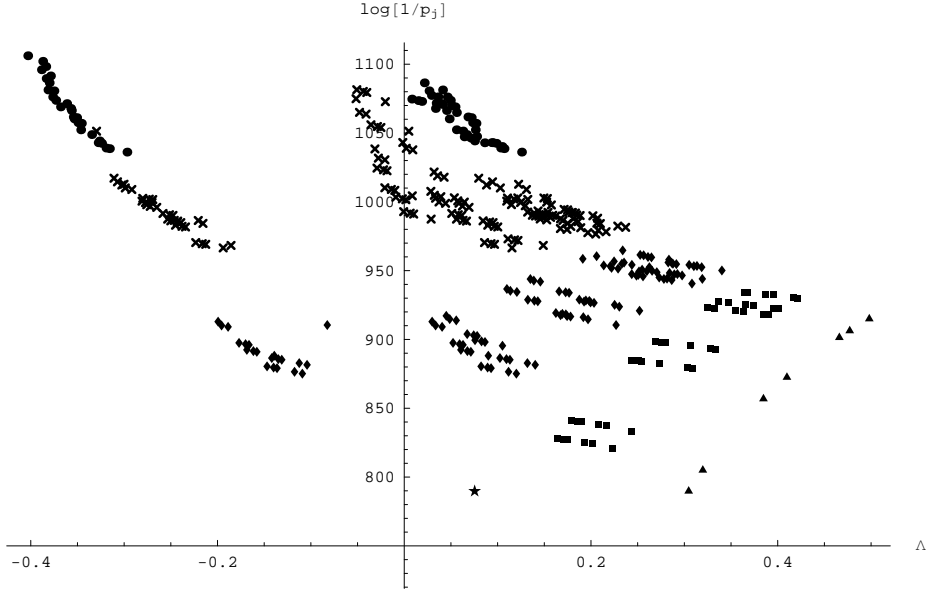}
\end{center}
\caption{ Plot of $\log_{10}(1/p_j)$ vs. $\Lambda_j$ for the BP
model with parameters given in (\ref{7D}).   The star marks the
dominant vacuum $\alpha_*$.  Triangles represent vacua in group 1,
squares in group 2, diamonds in groups 3 and 6, crosses in groups 4
and 7, and points in groups 5 and 8.} \label{7Dprobabilities}
\end{figure}

The bubble abundance factors $p_j$ for the 7-dimensional toy model
(\ref{7D}) are shown in Fig. \ref{7Dprobabilities}.  We plot
$\log_{10}(1/p_j)$ vs. $\Lambda_j$, so higher points in the figure
correspond to smaller bubble abundances.  The first thing one
notices is that there are several groups of points, marked by
triangles, boxes, etc.
The star marks the dominant vacuum $\alpha_*$.

In this particular example, the dominant site has coordinates
$(1,1,1,1,1,1,1)$.  There are $J=7$ ways to jump up one unit from this
site, to arrive at the seven points indicated by black triangles,
which we shall call group 1. The coordinates of these points are
$(2,1,1,1,1,1,1),~ (1,2,1,1,1,1,1),~..., ~(1,1,1,1,1,1,2)$.
Lower-energy vacua in this group have higher bubble abundances, in
accordance with Eqs.~(\ref{GammaI}),(\ref{ILambda}) of Section III.

The next group of states results from downward jumps out of vacua in
group 1 in all possible directions, excluding the jumps back to the
dominant site $\alpha_*$. We call it group 2. The number of states
in this group is $J(J-1)=42$.  Consider, for example, the subgroup
of states in group 2 coming from the downward transitions out of the
state $(2,1,1,1,1,1,1)$.  These states have coordinates
$(2,0,1,1,1,1,1),~...,~(2,1,1,1,1,1,0)$.  Since they originate from
the same single parent, the difference in their bubble abundances
comes from the difference in the instanton actions $I_{ij}$. This
effect is much milder for downward transitions than it is for the
upward ones.  That is why the spread in bubble abundances within the
subgroups of group 2 is much smaller than it is in group 1.

Further downward jumps replacing one of the $J-2=5$ remaining 1's by
a 0 give rise to group 3, consisting of $J(J-1)(J-2)/{2!}=105$
states having flux configurations with one $n=2$, four $n=1$ and two
$n=0$. Similarly, group 4 includes $J(J-1)(J-2)(J-3)/{3!}= 140$
states with one $n=2$, three $n=1$ and three $n=0$, and group 5
includes $J(J-1)(J-2)(J-3)(J-4)/4!= 105$ states with one $n=2$, two
$n=1$ and four $n=0$.  The factorial factors are included to avoid
double counting. For example, the site $(2,0,1,1,0,1,1)$ can be
reached by downward jumps from either $(2,1,1,1,0,1,1)$ or
$(2,0,1,1,1,1,1)$ and would be counted twice if we did not divide by
$2!$.

If a vacuum in group 2 has a coordinate jump from $n=2$ to $n=1$,
the resulting site is one of the daughter sites which can also be
reached by downward jumps from the dominant site.  These
negative-$\Lambda$ vacua have non-zero probabilities already at the
zeroth-order level and are not represented in the figure.

If a vacuum in group 3, 4 or 5 has a coordinate jump from $n=2$ to
$n=1$, the resulting sites are all terminal vacua (groups 6, 7 and
8, respectively).

We note that although the dominant vacuum $\alpha_*$ is one of the
low-energy vacua, there are many other recyclable vacua which have
lower $\Lambda$.  Recall that the dominant vacuum has the smallest,
in magnitude, sum of its transition rates down in each possible
direction (see Eq.'s (\ref{Ralpha}) and (\ref{q0})).  Each
transition rate depends exponentially on the value of $q_a$ and the
factor $r(x,y)/(n_a-1/2)^3$. From this factor and Fig.(\ref{7Drvsz})
we see that for $\Lambda/\Delta\Lambda<1$ (this is the case for
$\alpha_*$) any jump from an $n=2$ flux quanta will be less
suppressed than a jump in the same direction from an $n=1$ flux
quanta.  Thus we are not surprised that states which contain a flux
quanta of $n=2$ are \emph{not} dominant sites despite having smaller
$\Lambda$ than $\alpha_*$. Since each transition rate is
exponentially dependent on the tunneling exponent, typically the
largest (in magnitude) transition rate will dominate the sum in
Eq.(\ref{Ralpha}). Thus, essentially for a vacuum to be the dominant
state its largest (in magnitude) transition rate should be smaller
that the largest transition rate of any other vacuum.


The distribution in Fig. \ref{7Dprobabilities} was obtained in the
first order of perturbation theory, which includes only the vacua
which can be reached by a single upward jump from the dominant site
$\alpha_*$, followed by some downward jumps. If higher orders were
included, we would see additional groups of vacua, reachable only
with two or more upward jumps. These vacua would have much smaller
bubble abundances than those already in the figure.

The distribution $p_j$ in Fig. \ref{7Dprobabilities} spans more than
300 orders of magnitude. It differs dramatically from the flat
distribution (\ref{flat}) suggested by the heuristic argument in the
Introduction. Many vacua with close values of $\Lambda_j$ have very
different abundances $p_j$. The reason is that despite their
closeness in $\Lambda$, such vacua are located far away from one
another in the BP grid, and the paths leading to them from the
dominant vacuum $\alpha_*$ are characterized by exponentially
different transition rates. Even the vacua resulting from tunneling
out of the same site typically have very different abundances, due
to the exponential dependence of the tunneling rates on $q_a$.

\begin{figure}
\begin{center}
\leavevmode\epsfxsize=5in\epsfbox{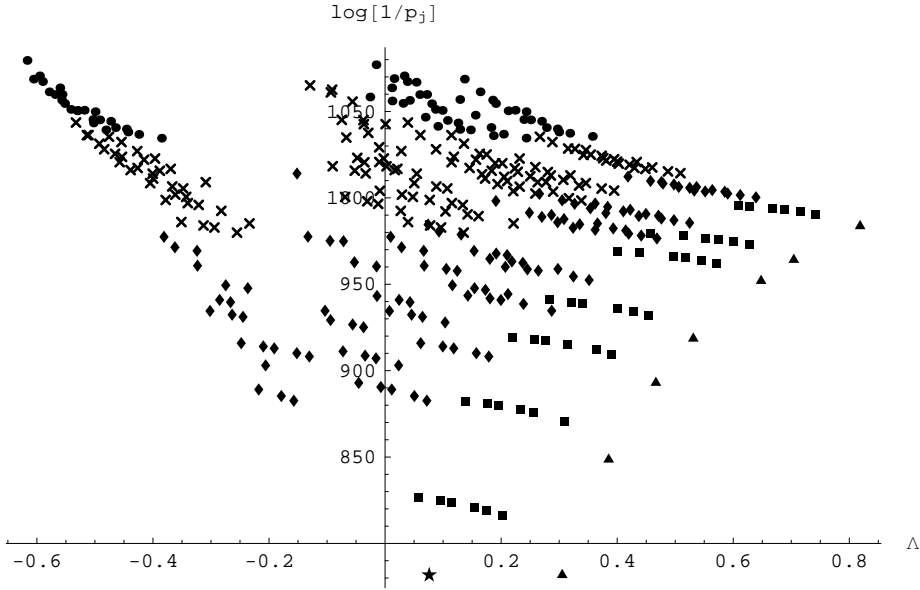}
\end{center}
\caption{ Plot of $\log_{10}(1/p_j)$ vs. $\Lambda_j$ for the BP
model with parameters (\ref{7D'}).  The star marks the dominant
vacuum $\alpha_*$.  Different groups of vacua are represented by the
same symbols as in Fig.5.} \label{7Dprobset2}
\end{figure}

Fig. \ref{7Dprobset2} shows the distribution of bubble abundances
for another $J=7$ model, with a different set of parameters: \beq
{q_i}= {0.6175,~0.3909,~0.6472,~0.5508,~0.5103,~0.7036,
0.4541};~~~~~~~\Lambda_{bare}=-1.033. \label{7D'} \eeq  In this
case, there is more scatter in the values of $q_a$, and the groups
of vacua are somewhat less pronounced. However, the staggered nature
of the distribution is still apparent.

\section{Discussion}

In this paper we have used the prescription of Refs.
\cite{GSPVW,ELM} to determine the bubble abundances $p_j$ in the BP
model. We found that the resulting distribution is very irregular,
with values of $p_j$ soaring and plummeting wildly as $\Lambda_j$
changes from one value to the next. This distribution is drastically
different from the flat distribution (\ref{flat}) which was used as
a basis for the successful anthropic prediction for $\Lambda$.

Apart from the bubble abundance factor $p_j$, the volume distribution
(\ref{PpZ}) includes the slow-roll expansion factor $Z_j$. In any
realistic model, bubble nucleation should be followed by a period of
slow-roll inflation, at least in some bubble types. The expansion
factor $Z_j$ is, of course, model-dependent, but there is no reason to
expect that it will somehow compensate for the wild swings in the
values of $p_j$ as we go from one value of $\Lambda_j$ to the next.

Another point to keep in mind is that, in a realistic setting, vacua
with different values of the fluxes $F_a$ may have different
low-energy physics, so the density of observers $n_{obs}(\Lambda)$
would also be very different. We should therefore focus on the
subset of vacua in the BP grid which differ only by the value of
$\Lambda$ and have essentially identical low-energy constants. Once
again, there seems to be no reason to expect any correlation between
these constants and the up and down swings in the bubble abundances.
We conclude that the staggered character of the distribution $P_j
\equiv P(\Lambda_j)$ is expected to persist, even in more realistic
versions of the model.

This conclusion is not limited to the BP model. It is likely to
arise in any landscape scenario, where a dense spectrum of
low-energy constants is generated from a wide distribution of states
in the parameter space of the fundamental theory. Vacua with nearly
identical values of $\Lambda$ may then come from widely separated
parts of the landscape and may have very different bubble abundances
and volume fractions.

Given the staggered character of the volume distribution, what kind
of prediction can we expect for the observed value of $\Lambda$? The
answer depends on the number ${\cal N}_A$ of possible vacua with
$\Lambda_j$ within the anthropic range (\ref{DeltaLambdaA}),
$\Delta\Lambda_A \sim 10^{-118}$. (We count only vacua in which all
low-energy constants other than $\Lambda$ have nearly the same
values as in our vacuum.)

Suppose the volume factors in the distribution $P_j$ span $K$ orders
of magnitude. ($K\sim 300$ in our numerical example in Section V.)
We can divide all vacua into, say, $10K$ bins, such that the values
of $P_j$ in each bin differ by no more than $10\%$. Suppose now that
there are ${\cal N}_A \ll 10K$ vacua in the anthropic range
$\Delta\Lambda_A$. We can then expect that most of these vacua will
be characterized by vastly different volume factors $P_j$, so that
the entire range will be dominated by one or few values of
$\Lambda_j$ having much higher volume fractions than the rest.

Moreover, there is a high likelihood of finding still greater volume
fractions if we go somewhat beyond the anthropic range - simply
because we would then search in a wider interval of $\Lambda$.  We
could, for example, find that a vacuum with $\Lambda_1 \sim
10^{-114}\sim 10^6\Lambda_0$ has a volume fraction 200 orders of
magnitude greater than all other vacua in the range
$0<\Lambda\lesssim \Lambda_1$. Galaxy formation is strongly
suppressed in this vacuum: the fraction of matter that ends up in
galaxies is only $f_G(\Lambda_1)\sim 10^{-110}$.  However, this
suppression is more than compensated for by the enhancement in the
volume fraction.

If this were the typical situation, most observers would find
themselves in rare, isolated galaxies, surrounded by nearly empty
space, all the way to the horizon.  This is clearly not what we
observe. The dominant value could by chance be very close to
$\Lambda=0$, but if such an ``accident'' is required to explain the
data, the anthropic model loses much of its appeal.

In the opposite limit, ${\cal N}_A\gg 10K$, the number of vacua in the
anthropic interval $\Delta\Lambda_A$ is so large that they may scan the
entire range of $P_j$ many times. Then, it is conceivable that the
distribution will become smooth after averaging over some suitable
scale $\delta\Lambda$. If $\delta\Lambda$ can be chosen much smaller
than $\Delta\Lambda_A$, then it is possible that the effective,
averaged distribution will be flat, as suggested by the heuristic
argument in the Introduction.  The successful prediction for $\Lambda$
would then be unaffected.\footnote{Joe Polchinski has informed us that
a similar argument, indicating that the anthropic explanation for the
observed $\Lambda$ requires a large number of vacua in the anthropic
range, was suggested to him by Paul Steinhardt.}

The above argument is somewhat simplistic, as it assumes that the
vacua in the BP grid are more or less randomly distributed between the
$10K$ bins, with roughly the same number of vacua in each bin. Such an
``equipartition'' is not likely to apply to the most abundant vacua,
but it may hold for the vacua in the mid-range of $P_j$. Finding the
conditions under which equipartition applies would require a
statistical analysis that goes beyond the scope of the present paper.

In summary, it appears that the staggered volume distribution
resulting from the BP model is in conflict with observations, unless
it yields a huge number of vacua in the anthropic range of
$\Lambda$. Counting only vacua which have nearly the same low-energy
physics as ours, we should have much more than $10K$; hence, the
total number of vacua should be many orders of magnitude greater.
The large number of vacua in the anthropic range is only a necessary
condition for the distribution $P_j$ to average out to the flat
distribution (\ref{flat}). Further analysis will be needed to find
whether or not this actually happens, and if so, then under what
conditions.  It would also be interesting to analyze other simple
models of the landscape, such as the ``predictive landscape'' of
Arkani-Hamed, Dimopoulos and Kachru \cite{AHDK}, and see what
similarities and differences they have compared to the BP model.

Throughout this paper we assumed that the brane charges $q_a$ are
not particularly small. This assumption may be violated in certain
parts of the landscape, e.g., in the vicinity of conifold points,
resulting in a much denser spectrum of vacua
\cite{FMSW,AshokDouglas,DenefDouglas,Giryavets}. Infinite
accumulations of vacua may occur near certain attractor points
\cite{DVattractor,Gia}.  Implications of these effects for the
probability distribution on the landscape remain to be explored.

\section{Acknowledgements}

 We are grateful to Michael Douglas, Gia Dvali, Jaume Garriga, Ken
Olum and Joseph Polchinski for useful comments and discussions.  This
work was supported in part by the National Science Foundation.

\section{Appendix}

\begin{figure}
\begin{center}
\leavevmode\epsfxsize=5in\epsfbox{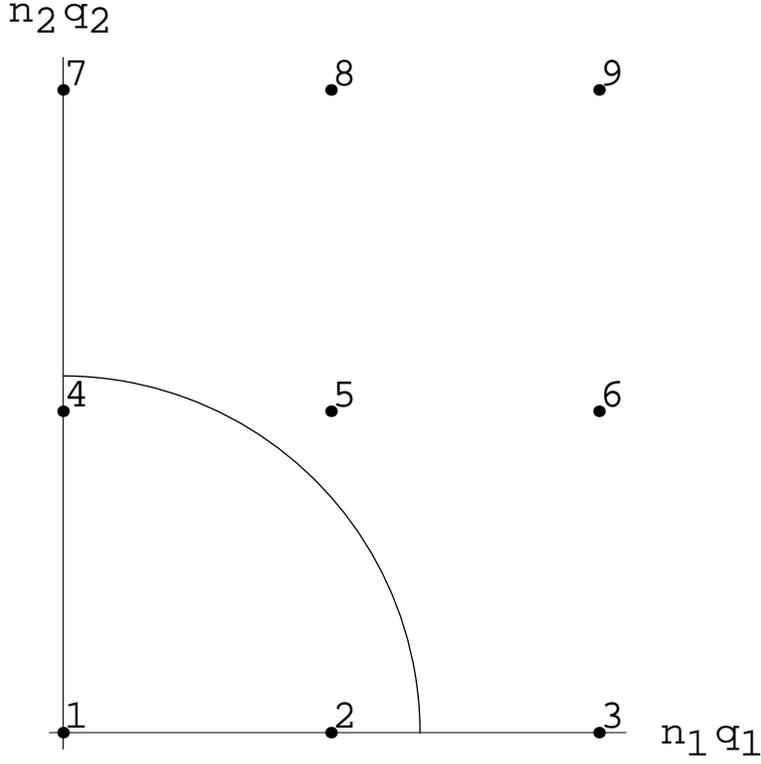}
\end{center}
\caption{The arrangement of vacua for a $J=2$, $N=2$ BP
grid } \label{2DanalyticBP}
\end{figure}

We will illustrate our perturbative method of calculation on a very
basic BP model, which can be solved analytically. We consider $9$
vacua arranged in a 2-D grid and labeled as indicated in Fig.
\ref{2DanalyticBP}. There are three terminal vacua labeled 1, 2, 4,
and six non-terminal vacua, 3, 5, 6, 7, 8, 9 in this model. We allow
upward and downward transitions between nearest neighbor pairs, with
transitions from non-terminal to terminal states allowed, but no
transitions may take place from a terminal state.  For simplicity,
we disregard the vacua in the quadrants where $n_1<0$ and/or $n_2<0$
and assume that the set of vacua in Fig.(\ref{2DanalyticBP}) is all
there is.

The evolution equations for the set of \emph{non-terminal} vacua is
\be
\frac{d\mathbf{f}}{dt}=\mathbf{R}\mathbf{f},
\ee
where the vector $\mathbf{f} \equiv \{f_3, f_5, f_6, f_7, f_8, f_9\}$.
Assuming that upward transition rates are far more suppressed
than downward transition rates, we represent the transition matrix as
\be
\mathbf{R}=\mathbf{R^{(0)}}+
\mathbf{R^{(1)}},
\ee
where
\be
\mathbf{R^{(0)}}=
\left(%
\begin{array}{cccccc}
  -D_3 & 0 & \kappa_{36} & 0 & 0 & 0 \\
  0 & -D_5& \kappa_{56} & 0 & \kappa_{58}& 0 \\
0& 0& -D_6& 0 & 0& \kappa_{69} \\
  0 & 0 & 0& -D_7 & \kappa_{78}& 0 \\
    0 & 0 & 0&  0& -D_8 & \kappa_{89}\\
      0 & 0 & 0& 0& 0&-D_9 \\
\end{array}%
\right)
\label{R0}
\ee
and
\be
\mathbf{R^{(1)}}=
\left(%
\begin{array}{cccccc}
  -U_3 & 0 & 0& 0 & 0 & 0 \\
  0 & -U_5& 0 & 0 & 0& 0 \\
\kappa_{63}& \kappa_{65}& -U_6& 0 & 0& 0 \\
  0 & 0 & 0& -U_7 & 0& 0 \\
    0 & \kappa_{85} & 0&  \kappa_{87}& -U_8 & 0\\
      0 & 0 & \kappa_{96}& 0& \kappa_{98}&-U_9 \\
\end{array}%
\right)
\label{R1}
\ee
and we have defined
\beqa
D_\alpha \equiv
\sum_{j<\alpha}\kappa_{j\alpha},\\ U_\alpha \equiv
\sum_{j>\alpha}\kappa_{j\alpha}.
\eeqa
$ D_\alpha$ and $U_\alpha$ represent, respectively, the sum of the
downward and upward transition rates from vacuum $\alpha$.

In our toy model we will assume that vacuum $5$ has the smallest (in
magnitude) sum of downward transition rates, and therefore $-q^{(0)} =
-D_5$ is the zero'th order dominant eigenvalue.  By inspection, we see
that the corresponding eigenvector is $\mathbf{s^{(0)}} \equiv \{0, 1, 0,
0, 0, 0\}$.

We now need to calculate the first order correction to this
eigenvector, $\mathbf{s^{(1)}}$. Substituting (\ref{R0}),(\ref{R1}) in
Eq.~(\ref{firstorder}), we find
\be
\left(%
\begin{array}{cccccc}
  q^{(0)}-D_3 & 0 & \kappa_{36} & 0 & 0 & 0 \\
  0 & 0& \kappa_{56} & 0 & \kappa_{58}& 0 \\
0& 0& q^{(0)}-D_6& 0 & 0& \kappa_{69} \\
  0 & 0 & 0&q^{(0)} -D_7 & \kappa_{78}& 0 \\
    0 & 0 & 0&  0& q^{(0)}-D_8 & \kappa_{89}\\
      0 & 0 & 0& 0& 0&q^{(0)}-D_9 \\
\end{array}%
\right)\mathbf{s^{(1)}} =\left(%
\begin{array}{c}
  0 \\
   -q^{(1)}+U_5 \\
-\kappa_{65}\\
  0 \\
   -\kappa_{85}\\
      0  \\
\end{array}%
\right)
\label{matrixinvert}
\ee
Note that the only equation in this set that depends on the first
order correction to the eigenvalue is also the equation that needs to
be dropped from our system, since it has a zero diagonal element -
this causes the matrix on the right-hand side of
(\ref{matrixinvert}) to have a zero determinant, which
renders it uninvertible.

This drop in the number of independent equations is replenished by
including the constraint equation (\ref{constraint}); the resulting
set of equations is
\be \left(%
\begin{array}{cccccc}
  q^{(0)}-D_3 & 0 & \kappa_{36} & 0 & 0 & 0 \\
  0 & 1& 0 & 0 & 0& 0 \\
0& 0& q^{(0)}-D_6& 0 & 0& \kappa_{69} \\
  0 & 0 & 0&q^{(0)} -D_7 & \kappa_{78}& 0 \\
    0 & 0 & 0&  0& q^{(0)}-D_8 & \kappa_{89}\\
      0 & 0 & 0& 0& 0&q^{(0)}-D_9 \\
\end{array}%
\right)\mathbf{s^{(1)}} =\left(%
\begin{array}{c}
  0 \\
  0 \\
-\kappa_{65}\\
  0 \\
   -\kappa_{85}\\
      0  \\
\end{array}%
\right)\ee

The solution is readily determined, and we obtain
\be
\mathbf{s^{(1)}} = \left(%
\begin{array}{c}
 \frac{\kappa_{36}}{D_3-D_5} \frac{\kappa_{65}} {D_6-D_5}\\
  0 \\
\frac{\kappa_{65}}{D_6- D_5}\\
 \frac{\kappa_{78}}{D_7-D_5} \frac{\kappa_{85}} {D_8-D_5} \\
  \frac{\kappa_{85}} {D_8-D_5}\\
      0  \\
\end{array}%
\right) \ee This can now be used in Eq.(\ref{pJaume}) to determine
the bubble abundances. For example, comparing the bubble abundances
in vacua 3 and 7, we find \be \frac{p_3}{p_7}= {H_6^q
\kappa_{36}\kappa_{65}(D_8 - D_5)
  \over{H_8^q \kappa_{78}\kappa_{85}(D_6 - D_5)}} \ee


\end{document}